\documentclass[superscriptaddress,preprint]{revtex4} % Define the document class

\usepackage{graphicx}                             % Package for graphics

\newcommand{\ket}[1]{\ensuremath{|#1\rangle}}
\newcommand{\bra}[1]{\ensuremath{\langle#1|}}

\begin{document}

\title{Selective darkening of degenerate transitions demonstrated with two superconducting quantum bits}

\author{P.C.~de Groot}
\email{p.c.degroot@tudelft.nl}
\affiliation{Kavli Institute of Nanoscience, Delft University of Technology, P.O. Box 5046, 2600 GA Delft, The Netherlands}

\author{J.~Lisenfeld}
\affiliation{Kavli Institute of Nanoscience, Delft University of Technology, P.O. Box 5046, 2600 GA Delft, The Netherlands}
\affiliation{Physikalisches Institut and DFG Center for Functional Nanostructures (CFN), Karlsruhe Institute of Technology, D-76131 Karlsruhe, Germany}

\author{R.N.~Schouten}
\affiliation{Kavli Institute of Nanoscience, Delft University of Technology, P.O. Box 5046, 2600 GA Delft, The Netherlands}

\author{S.~Ashhab}
\affiliation{Advanced Science Institute, The Institute of Physical and Chemical Research (RIKEN), Wako-shi, Saitama 351-0198, Japan}
\affiliation{Department of Physics, University of Michigan, Ann Arbor, Michigan 48109-1040, USA}

\author{A.~Lupa\c{s}cu}
\affiliation{Institute for Quantum Computing, University of Waterloo, N2L 5G7 Waterloo, Canada}

\author{C.J.P.M.~Harmans}
\affiliation{Kavli Institute of Nanoscience, Delft University of Technology, P.O. Box 5046, 2600 GA Delft, The Netherlands}

\author{J.E.~Mooij}
\affiliation{Kavli Institute of Nanoscience, Delft University of Technology, P.O. Box 5046, 2600 GA Delft, The Netherlands}

\date{\today}

\maketitle

%%%%%%%%%%%%% Introduction

\textbf{
Controlled manipulation of quantum states is central to studying natural and artificial quantum systems.
If a quantum system consists of interacting sub-units, the nature of the coupling may lead to quantum levels with degenerate energy differences.
This degeneracy makes frequency-selective quantum operations impossible.
For the prominent group of transversely coupled two-level systems, i.e. qubits, we introduce a method to selectively suppress one transition of a degenerate pair while coherently exciting the other, effectively creating artificial selection rules.
It requires driving two qubits simultaneously with the same frequency and specified relative amplitude and phase.
We demonstrate our method on a pair of superconducting flux qubits \cite{mooij:99}.
It can directly be applied to the other superconducting qubits \cite{nakamura:99,vion:02,martinis:02,schreier:08,manucharyan:09}, and to any other qubit type that allows for individual driving.
Our results provide a single-pulse controlled-NOT gate for the class of transversely coupled qubits.
}

\vspace{10mm}

%%%%%%%%%%%%%%%%% Previous experiments

  Transverse coupling implies an interaction that is transverse to the eigenstates of the uncoupled systems. For this type of coupling the energy splitting of one qubit does not depend on the state of the other. This property is appealing because it means that in the absence of driving the system essentially behaves as a set of uncoupled qubits.
  The coupling primarily manifests itself when the system is driven and hence can be regarded as AC-tunable \cite{paraoanu:06}.
  A coupling that is not purely transverse leads to a spectroscopic splitting of the transitions. Although this splitting enables simple resonant driving for all operations \cite{linden:98,plantenberg:07}, in practice it requires refocussing schemes to compensate for the continuously evolving phases \cite{jones:99}.
  The price to pay for the advantage of transverse coupling is obvious; the degeneracy prohibits schemes for selective excitation that rely on a frequency splitting.
  Previous experiments used either additional coupling elements \cite{niskanen:07}, extra modes \cite{sillanpaa:07}, or shifted levels into and out of resonance by DC \cite{mcdermott:05} or strong AC fields \cite{rigetti:05,majer:07}.
  Note that level shifting can imply passing through conditions of low coherence \cite{simmonds:04}, or passing resonances with other qubits.
  In contrast, our method works for simple direct coupling as well as for systems with additional coupling elements, such as harmonic oscillators, as long as the effective coupling is transverse. It uses only a single pulse of a single frequency and does not require (dynamical) shifting of the levels.

%%%%%%%%%%% method

  We consider the class of systems of transversely coupled qubits, described with the Hamiltonian
  \begin{equation}
    H = - \frac{1}{2} \left( \Delta_1 \sigma_z^1 + \Delta_2 \sigma_z^2 \right) + J \sigma_x^1 \sigma_x^2,
    \label{eq:H}
  \end{equation}
  where \( \Delta_i \) is the single-qubit energy splitting of qubit \( i \), \( \Delta_1~\neq~\Delta_2 \), \( J \) is the qubit-qubit coupling energy and \( \sigma_{x,y,z}^i \) are the Pauli spin matrices.
  This Hamiltonian describes many actively used quantum systems \cite{mooij:99,nakamura:99,vion:02,martinis:02,schreier:08,manucharyan:09}, and often applies for operation at a coherence sweet-spot \cite{vion:02,yoshihara:06,rigetti:05,liu:06,bertet:06}.
  The energy levels of this system are shown schematically in Fig. 1b. The arrows indicate the transitions of interest; the blue and red arrows describe the transitions of qubit 1 and 2, respectively. Both pairs are degenerate in frequency, which is typical for transverse coupling.

  Our method aims at the selective excitation of these transitions and is based on simultaneously driving the two qubits with a single resonant frequency, employing different amplitudes and phases.
  The driving is described with the Hamiltonian
    \begin{equation}
      H_{\rm drive} = \
          a_1 \cos(\omega t + \varphi_1) \sigma_x^1
        + a_2 \cos(\omega t + \varphi_2) \sigma_x^2,
      \label{eq:Hdrive}
    \end{equation}
where \( \omega \) is the driving frequency, and \( a_i \) and \( \varphi_i \) are the driving amplitude and phase for qubit \( i \).
The transition strength \( T_{k \leftrightarrow l} = \langle l | \tilde{H}_{\rm drive} | k \rangle \), with the driving Hamiltonian transformed to an appropriate rotating frame (see Supplementary Information), governs the transition rate and depends on both \( a_i \) and \( \varphi_i \).
  Figure 1d shows the normalized \( |\overline{T}| = |T|/(a_1+a_2) \) as a function of \( a_1/(a_1+a_2) \), for a fixed phase difference \( \varphi_2-\varphi_1 = 0 \).
    Clearly the two transitions of each qubit generally do \emph{not} have the same strength \( T \), despite their frequency degeneracy.
    In addition, for certain settings individual transitions are completely suppressed: the transition is darkened (black dotted lines in Fig. 1d).
    The darkened transitions provide the desired conditions where one of the two transitions can be excited individually, even though the driving field is resonant with both transitions.
    The difference in transition strength can be understood intuitively. As coupling leads to mixing of the single-qubit eigenstates, qubit 2 can be excited by driving qubit 1 with a frequency that is resonant with qubit 2. This indirect driving can enhance, counteract and even cancel direct driving of qubit 2. The effect differs for the two degenerate transitions, because the states involved are different superpositions of the single-qubit eigenstates (see Supplementary Information).
    For \( J \ll | \Delta_1 - \Delta_2 | \), and assuming \( \Delta_1 > \Delta_2 \), one readily finds the driving amplitude ratio
    \begin{equation}
        \frac{a_2}{a_1} = \frac{J}{\Delta_1-\Delta_2},
      \label{eq:Tij}
    \end{equation}
    which yields \( T_{00 \leftrightarrow 01} = 0 \) for \( \varphi_2 - \varphi_1 = 0 \) and \( T_{10 \leftrightarrow 11} = 0 \) for \( \varphi_2 - \varphi_1 = \pi \). These are the transitions of qubit 2. For the transitions of qubit 1, i.e. \( 00 \leftrightarrow 10 \) and \( 01 \leftrightarrow 11 \), the amplitude ratio is simply inverted.
    Expressions for arbitrary \( J \) are given in the Supplementary Information.

%%%%%%%%%%%%% flux qubits

To experimentally demonstrate this method we employ two coupled flux qubits \cite{mooij:99}, each consisting of a superconducting loop interrupted by four Josephson tunnel junctions.
  When biased with a magnetic flux close to half a flux quantum \( \Phi_0 \), the two states of each qubit are clockwise and anti-clockwise persistent-current states. These currents \( I_p \) produce opposite magnetic fields, which  provides the coupling for the two qubits.
  Two independent AC-operated SQUID magnetometers are used to simultaneously read out the states of the qubits \cite{lupascu:07,groot:10}. These are switching-type detectors, where the switching probability \( P_{sw} \) is a measure for the magnetic field.
  At a bias of \( \frac{\Phi_0}{2} \) the system is described by the Hamiltonian of Eq. 1.
  Here the eigenstates of each qubit are symmetric and anti-symmetric superpositions of the two persistent-current states, with level separation \( \Delta \).
   The device is depicted in Fig. 1a.
         The qubits are characterised by the persistent currents \( I_{p,1}=355~\mathrm{nA}\) and \( I_{p,2}=460~\mathrm{nA}\) and the energy splittings \( \Delta_{1}/h=7.88~\mathrm{GHz} \) and \( \Delta_{2}/h=4.89~\mathrm{GHz} \). The qubit-qubit coupling strength is \( 2J/h = 410~\mathrm{MHz} \).

%%%%%%%%%%%%%% implementation driving

  For our fabricated quantum objects the spatial locations are well-defined, and the individual control of amplitude and phase for each qubit according to Eq. 2 can be easily achieved using local magnetic fields.
  We employ two on-chip antennas, indicated as \( A_1 \) and \( A_2 \) in Fig. 1a, both coupling to both qubits, with a stronger coupling to the closer one. Driving the two qubits from both antennas is described with
    \begin{eqnarray}
      H_{\rm drive} = \
          A_1 \cos(\omega t + \phi_1)
          ( m_{11} \sigma_x^1 + m_{12} \sigma_x^2) \nonumber \\
        + A_2 \cos(\omega t + \phi_2)
          ( m_{21} \sigma_x^1 + m_{22} \sigma_x^2),
      \label{eq:Hdrivet}
    \end{eqnarray}
    where \( A_j \) and \( \phi_j \) are the driving amplitude and phase for antenna \( j \) and \( m_{ji} \) is the coupling of antenna \( j \) to qubit \( i \).
    Note that \emph{any} combination of \( a_i \) and \( \varphi_i \) in Eq. 2 can be achieved with the proper choice of \( A_j \) and \( \phi_j \). For this device \( m_{12}/m_{11}=0.32 \), \( m_{21}/m_{22}=0.33 \) and \( m_{11}=m_{22} \).

%%%%%%%%%%%%%%%%%%% Results1: driving from a single antenna

For the experimental demonstration we choose to focus on the degenerate transitions of qubit 2. We first show that, if the qubits are driven from a \emph{single antenna}, the two degenerate transitions exhibit a different Rabi frequency.
We apply two pulses on antenna 1: the first pulse is resonant with qubit 1, the second pulse is resonant with qubit 2. Figure 1c shows a schematic of the pulse-sequence; note that here \( A_2=0 \). The experiment is repeated for varying durations \( \tau_1 \) and \( \tau_2 \) of pulses 1 and 2.
  The switching probability \( P_{sw,1} \) of detector 1 is depicted in Fig. 2a, showing a few Rabi oscillation periods as a function of the pulse duration \( \tau_1 \). Varying \( \tau_2 \) does not lead to oscillations of qubit 1, as pulse 2 is non-resonant, and only relaxation is observed.
  The oscillations of qubit 2, induced by the second pulse, are visible in \( P_{sw,2} \) (Fig. 2b). Here we distinguish two oscillation frequencies. Along the white solid line, where qubit 1 is prepared in the excited state, qubit 2 oscillates with a Rabi frequency of \( 85~\mathrm{MHz} \). For qubit 1 prepared in the ground state, along the white dashed line, the Rabi frequency \( f=31~\mathrm{MHz} \) of qubit 2 is lower.
  For qubit 1 in a superposition of the ground and excited states, qubit 2 shows a beating pattern of both oscillations.

  A more detailed analysis allows us to unravel the two frequencies of Fig. 2b and determine which levels are participating in each of the oscillations. We extract the level occupations \( Q_{00}, Q_{01}, Q_{10} \) and \( Q_{11} \) from the individual switching probabilities (see Supplementary Information). The result is shown in Figures 2c-f. After an odd number of \( \pi \)-rotations of qubit 1, there are only oscillations between states 10 and 11, not for states 00 and 01. After an even number of \( \pi \)-rotations of qubit 1 the situation is reversed; now the states 00 and 01 oscillate. The two oscillation frequencies are clearly linked to the two different transitions.

%%%%%%%%%%%%%%%%%% Results2: driving from two antennas

  To demonstrate the \emph{tunability of the transition strengths} we drive both antennas simultaneously, using the same frequency and controlling independently the amplitudes  \( A_1, A_2 \) and phases \( \phi_1, \phi_2 \). In this 2-pulse experiment (Fig. 1c), the first pulse prepares qubit 1 with a \( \pi/2 \)-rotation and the duration \( \tau_2 \) of the second pulse is varied. Since qubit 1 is in a superposition state, both Rabi frequencies are present in the dynamics of qubit 2.
  In Figures 3a-c we show the Fourier transform for the measured oscillations. Each graph is measured with a different amplitude ratio \( A_1/A_2 \), with fixed phase \( \phi_1=0 \) and varying \( \phi_2 \).
  Figure 3a, with \( A_1/A_2=1.3 \), shows a typical result for an arbitrary amplitude-ratio; the Rabi oscillation frequencies of both transitions clearly depend on \( \phi_2 - \phi_1 \), but nowhere a transition is darkened. Note the occurrence of equal Rabi frequencies for two phase conditions, as denoted by Y. For \( A_1/A_2=2.5 \) in Figure 3b we observe that for \( \phi_2-\phi_1 \approx \pi \) (indicated by \( X_0 \)) the transition \( 10 \leftrightarrow 11 \) is fully darkened, while the \( 00 \leftrightarrow 01 \) transition shows a non-zero oscillation frequency. In Fig. 3c with \( A_1/A_2=6.3 \) the situation is reversed, with the \( 00 \leftrightarrow 01 \) transition being suppressed (denoted by \( X_1 \)). This clearly demonstrates our method, as we selectively excite one of two transitions, despite their frequency degeneracy.
  Calculations of \( T \) are in good agreement with the experimental results, provided we allow for different transmissions of amplitudes and phases of the antennas to the qubits, which we attribute to the influence of the detector circuits.

%%%%%%%%%%% Results3: selective excitation

To further investigate the special cases of equal Rabi frequencies (Y), and \emph{darkened transitions} (\( X_0 \), \( X_1 \)), we again vary the durations \( \tau_1 \) and \( \tau_2 \), using both antennas for the second pulse. The results should be compared with Fig. 2b.
  Figure 3d shows \( P_{sw,2} \) for driving conditions denoted by \( Y \) (left arrow): the oscillation frequency of qubit 2 does not depend on the state of qubit 1.
  For the conditions marked by \( X_0 \), we only observe oscillations of qubit 2 when qubit 1 is in the ground state, as shown in Fig. 3e.
  Similarly for the conditions marked by \( X_1 \), now we only observe oscillations of qubit 2 if qubit 1 is in the excited state (Fig. 3f).

  The demonstrated capability to selectively manipulate transition strengths in frequency-degenerate transitions has important applications. A \( \pi \)-pulse using condition \( X_1 \) or \( X_0 \) provides a 1-controlled and 0-controlled NOT gate, respectively. This enables certain systems, including the flux qubit used here, to be fully operated at the coherence-optimal point, without level shifting by either DC or strong AC signals. Note that the use of additional coupling elements is neither required nor prohibited. If additional coupling elements are used, our method can replace more complicated schemes. For conditions similar to \( Y \), taking care of the individual rotation angles, also single-qubit gates can be implemented.
  The controlled-NOT and single-qubit gates together form a universal set, implying that our method fulfills all requirements for constructing any single- or two-qubit gate.
  The method also scales to three or more qubits, provided that for a certain target pair the system can be reduced to Eq. 1.

%%%%%%%%%%%% Discussion and Conclusion

  In conclusion, we have introduced and experimentally demonstrated a method to control transition strengths by applying a non-uniform driving field. Darkened transitions are created and employed for the selective excitation of degenerate transitions. As this method improves the simplicity and coherence conditions for operations in a variety of quantum systems, the prospect of performing large-scale quantum algorithms is enhanced significantly.

%%%%%%%%%%%%%% Acknowledgements
\clearpage
\noindent
{\large{Acknowledgements}}

\noindent
We acknowledge L. M. K. Vandersypen, G. A. Steele, I. T. Vink, J. Baugh and the Delft Flux Qubit Team for help and discussions and A. van der Enden, R. G. Roeleveld, B. P. van Oossanen and the Nanofacility for technical and fabrication support. This work is supported by NanoNed, FOM, NSERC Discovery and EU projects EuroSQIP and CORNER.

%%%%%%%%%%%%%% Author Contribution
\vspace{10mm}
\noindent
{\large{Author Contribution}}

\noindent
P.C.G. and A.L. devised the method.
P.C.G. designed the experiment, and designed and fabricated the sample.
P.C.G. and J.L. carried out the experiments and analyzed the data.
S.A. analyzed the method theoretically.
R.N.S. developed and provided dedicated electronics.
P.C.G., J.E.M. and C.J.P.M.H. wrote the manuscript.
J.E.M. and C.J.P.M.H. supervised the project.

%%%%%%%%%%%%% Competing Financial Interests statement
\vspace{10mm}
\noindent
{\large{Competing financial interests}}

\noindent
The authors declare that they have no competing financial interests.

%%%%%%%%%%%%% References (replace with contents of .bbl)
\clearpage
%\bibliographystyle{naturemag}
%\bibliography{bibliography}

%%%%%%%%%%%%%%% Figure captions
\newpage
  \begin{figure}[htbp]
    \begin{center}
      \includegraphics{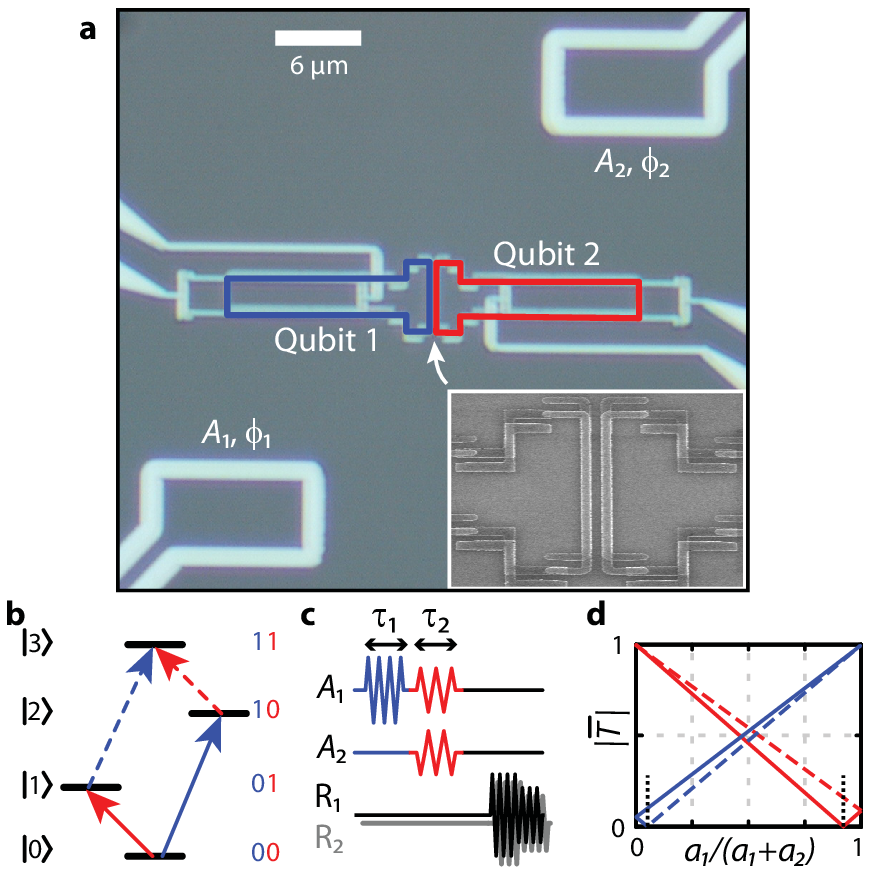}
    \end{center}
    \caption{
      \textbf{Coupled qubit system and transitions.}
      \textbf{a}, Optical micrograph of the sample, showing two flux qubits colored in blue and red.
         The inset shows part of each qubit loop, both containing four Josephson tunnel junctions.         %
         Overlapping the qubit loops, in light-grey, are the SQUID-based qubit-state detectors.
         In the top right and bottom left are the two antennas from which the qubits are driven.
      \textbf{b}, Energy level diagram of the coupled qubit system. Arrows of the same color indicate transitions of the same qubit and are degenerate in frequency.
      \textbf{c}, Pulse sequence used for the coherent excitation of the qubits. The first pulse is resonant with qubit 1. The second pulse, applied from both antennas simultaneously with independent amplitudes and phases, is resonant with qubit 2. After the second pulse the state of both qubits is read out.
      \textbf{d}, The normalized transition strengths of the four transitions in \textbf{b} as a function of the net driving amplitudes \( a_1/(a_1+a_2) \) for \( \varphi_2-\varphi_1=0 \). For \( \varphi_2-\varphi_1=\pi \) the dashed and solid lines are interchanged. The black dotted lines indicate the locations of the darkened transitions.
    }
    \label{fig:1}
  \end{figure}

\newpage
  \begin{figure}[htbp]
    \begin{center}
      \includegraphics{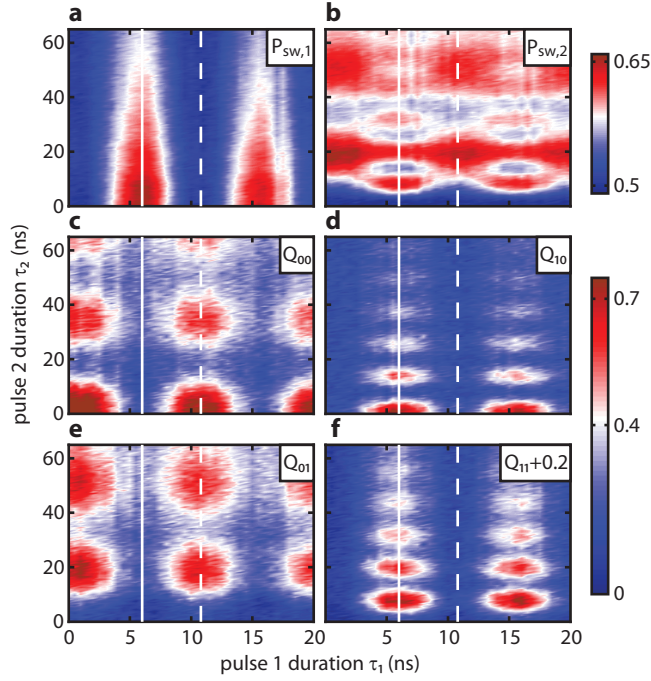}
    \end{center}
    \caption{
    \textbf{Driving from a single antenna.}
      Measurement of the state of the qubits, represented by switching probabilities \( P_{sw,1} \) and \( P_{sw,2} \), after applying a pulse of duration \( \tau_1 \) resonant with qubit 1, followed by a pulse of duration \( \tau_2 \) resonant with qubit 2.
      \textbf{a}, \( P_{sw,1} \), showing coherent oscillations of qubit 1 induced by pulse 1. The white solid and dashed lines indicate a \( \pi \)- and \( 2\pi \)-rotation respectively. For pulse 2, qubit 1 only shows relaxation.
      \textbf{b}, \( P_{sw,2} \), showing coherent oscillations induced by pulse 2. After an odd number of \( \pi \)-rotations on qubit 1, the oscillation frequency is higher than after an even number of \( \pi \)-rotations. For superposition states of qubit 1, a beating pattern of the two oscillations is observed.
      \textbf{c-f}, Level occupations \( Q \) of the four different levels. Note that a value of 0.2 has been added to \( Q_{11} \) to improve visibility.
    }
    \label{fig:2}
  \end{figure}

\newpage
  \begin{figure}[htbp]
    \begin{center}
      \includegraphics{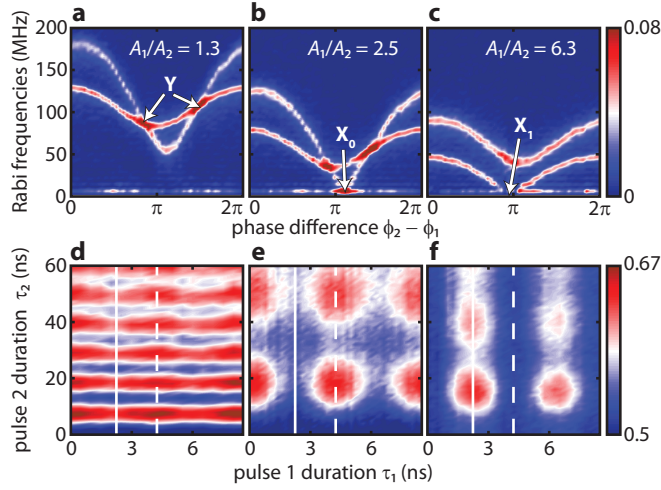}
    \end{center}
    \caption{
    \textbf{Transition strength tuning and darkened transitions.}
      \textbf{a-c}, Rabi frequency dependence on \( \phi_2-\phi_1 \) for three different amplitude-ratios.
      The color scale represents the Fourier component of \( P_{sw,2}(\tau_2) \). Qubit 1 is prepared with a \( \pi/2 \)-rotation.
      Markers \( X_0 \) and \( X_1 \) indicate the conditions for a darkened transition on \( 00 \leftrightarrow 01 \) and \( 10 \leftrightarrow 11 \) respectively.
      \textbf{d-f} \( P_{sw,2} \) versus the durations \( \tau_1 \) and \( \tau_2 \).
      The white solid and dashed lines indicate a \( \pi \)- and \( 2\pi \)-rotation of qubit 1, respectively.
      The driving conditions are as marked by Y left arrow (\textbf{d}), \( X_0 \) (\textbf{e}) and \( X_1 \) (\textbf{f}).
    }
    \label{fig:3}
  \end{figure}

\newpage
\renewcommand{\thefigure}{S\arabic{figure}}
\setcounter{figure}{0}
\setcounter{equation}{0}
{\Huge{Supplementary Information}}
\section{Conditions for darkened transition}

Let us take the Hamiltonian:
\begin{equation}
  H = -\frac{1}{2} \left(
      \Delta_1 \sigma_z^{1} + \Delta_2 \sigma_z^{2}
      \right) + J \sigma_x^{1} \sigma_x^{2},
\label{eq:Hamiltonian}
\end{equation}
with \( \Delta_1 \ne \Delta_2 \). Without loss of generality we take \( \Delta_1 > \Delta_2 \).
Additionally we have the driving term:
\begin{eqnarray}
H_{\rm drive}
& = &
  a_1 \cos (\omega t + \varphi_1) \sigma_x^{1}
+ a_2 \cos (\omega t + \varphi_2) \sigma_x^{2} \\
& = &
  \tilde{H}_{\rm drive}^{+} e^{ i \omega t}
+ \tilde{H}_{\rm drive}^{-} e^{-i \omega t},
\end{eqnarray}
where
\begin{equation}
  \tilde{H}_{\rm drive}^{\pm} =
      \frac{a_1}{2} e^{\pm i\varphi_1} \sigma_x^{1}
    + \frac{a_2}{2} e^{\pm i\varphi_2} \sigma_x^{2}.
\end{equation}
The Hamiltonian (Eq.~\ref{eq:Hamiltonian}) can be diagonalized without any approximations:
\begin{eqnarray}
  \ket{0} & = & \cos\theta_1 \ket{00} - \sin\theta_1
  \ket{11}
  \\
  \ket{1} & = & \cos\theta_2 \ket{01} - \sin\theta_2
  \ket{10}
  \\
  \ket{2} & = & \cos\theta_2 \ket{10} + \sin\theta_2
  \ket{01}
  \\
  \ket{3} & = & \cos\theta_1 \ket{11} + \sin\theta_1
  \ket{00},
\end{eqnarray}
where
\begin{eqnarray}
\tan 2\theta_1 & = & \frac{2J}{\Delta_1+\Delta_2}
\\
\tan 2\theta_2 & = & \frac{2J}{\Delta_1-\Delta_2}.
\end{eqnarray}
To calculate the transition strength \( T_{k \leftrightarrow l} = \bra{l} \tilde{H}_{\rm drive} \ket{k} \), we ignore the counter-rotating fields, which results in \( \tilde{H}_{\rm drive} = \tilde{H}_{\rm drive}^+ \) for \( k<l \) and \( \tilde{H}_{\rm drive} = \tilde{H}_{\rm drive}^- \) for \( k>l \). We find that
\begin{eqnarray}
\bra{1} \tilde{H}_{\rm drive} \ket{0} & = &
- \frac{a_1}{2} e^{i\varphi_1} \left\{
\cos\theta_1 \sin\theta_2 +
\sin\theta_1 \cos\theta_2 \right\} +
\frac{a_2}{2} e^{i\varphi_2} \left\{
\cos\theta_1 \cos\theta_2 +
\sin\theta_1 \sin\theta_2 \right\} \nonumber
\\
& = &
- \frac{a_1}{2} e^{i\varphi_1} \sin(\theta_1+\theta_2)
+ \frac{a_2}{2} e^{i\varphi_2} \cos(\theta_2-\theta_1)
\\
\bra{3} \tilde{H}_{\rm drive} \ket{2} & = &
+ \frac{a_1}{2} e^{i\varphi_1} \sin(\theta_1+\theta_2)
+ \frac{a_2}{2} e^{i\varphi_2} \cos(\theta_2-\theta_1),
\end{eqnarray}
and
\begin{eqnarray}
\bra{2} \tilde{H}_{\rm drive} \ket{0} & = &
\frac{a_1}{2} e^{i\varphi_1} \left\{
\cos\theta_1 \cos\theta_2 -
\sin\theta_1 \sin\theta_2 \right\} +
\frac{a_2}{2} e^{i\varphi_2} \left\{
\cos\theta_1 \sin\theta_2 -
\sin\theta_1 \cos\theta_2 \right\} \nonumber
\\
& = &
\frac{a_1}{2} e^{i\varphi_1} \cos(\theta_1+\theta_2) +
\frac{a_2}{2} e^{i\varphi_2} \sin(\theta_2-\theta_1)
\\
\bra{3} \tilde{H}_{\rm drive} \ket{1} & = &
\frac{a_1}{2} e^{i\varphi_1} \cos(\theta_1+\theta_2) -
\frac{a_2}{2} e^{i\varphi_2} \sin(\theta_2-\theta_1).
\end{eqnarray}
We therefore find that the required ratio of driving amplitudes to
drive only one of the two transitions of qubit 2 is given by
\begin{equation}
\frac{a_2}{a_1} =
\frac{\sin(\theta_1+\theta_2)}{\cos(\theta_2-\theta_1)},
\end{equation}
with $\varphi_2-\varphi_1=0$ for suppressing the \( 00 \leftrightarrow 01 \) transition, and $\varphi_2-\varphi_1=\pi$
for suppressing the \( 10 \leftrightarrow 11 \) transition. For driving the transitions of qubit 1, the ratio
is
\begin{equation}
\frac{a_2}{a_1} =
\frac{\cos(\theta_1+\theta_2)}{\sin(\theta_2-\theta_1)},
\end{equation}
with $\varphi_2-\varphi_1=\pi$ for suppressing the \( 00 \leftrightarrow 10 \) transition, and $\varphi_2-\varphi_1=0$
for suppressing the \( 01 \leftrightarrow 11 \) transition.

The resonance frequencies are given by:
\begin{eqnarray}
E_1-E_0 = E_3-E_2 & = & \frac{1}{2} \left(
\sqrt{4J^2+(\Delta_1+\Delta_2)^2} -
\sqrt{4J^2+(\Delta_1-\Delta_2)^2} \right)
\\
E_2-E_0 = E_3-E_1 & = & \frac{1}{2} \left(
\sqrt{4J^2+(\Delta_1+\Delta_2)^2} +
\sqrt{4J^2+(\Delta_1-\Delta_2)^2} \right).
\end{eqnarray}

If we now make the assumption that \( J \ll |\Delta_1 - \Delta_2| \), then to lowest order approximation \( \theta_1 = 0 \) and \( \theta_2 = J/|\Delta_1-\Delta_2| \). The eigenvectors reduce to
\begin{eqnarray}
  \ket{0} & = & \ket{00}
  \\
  \ket{1} & = & \ket{01} - \frac{J}{\Delta_1-\Delta_2} \ket{10}
  \\
  \ket{2} & = & \ket{10} + \frac{J}{\Delta_1-\Delta_2} \ket{01}
  \\
  \ket{3} & = & \ket{11},
  \label{eq:evsmallJ}
\end{eqnarray}
and \( E_1 - E_0 = E_3 - E_2 = \Delta_2 \), \( E_2 - E_0 = E_3 - E_1 = \Delta_1 \). Following the same procedure as above we find for the darkened transitions
\begin{equation}
  \frac{a_1}{a_2} = \frac{J}{\Delta_1-\Delta_2}
  \label{eq:sJq2}
\end{equation}
and
\begin{equation}
  \frac{a_2}{a_1} = \frac{J}{\Delta_1-\Delta_2}
  \label{eq:sJq1}
\end{equation}
for suppressing the transitions of qubit 1 and 2, respectively. The phase conditions are the same as above.

\section{Extracting level occupations from detector \\ switching-probabilities}
The detectors are click-type detectors; they either switch or do not switch away from the initial state, depending on the sensed magnetic field. Their fidelities are lower than 1, and the base levels for the switching rates are free to choose. Here we present a procedure to extract the level occupations \( Q_{i} =  |\langle i | \psi \rangle|^2 \), with \( i={00, 01, 10, 11} \) and \( \ket{\psi} \) the quantum state of the system.

Each individual measurement event can have one of four possible outcomes: neither detectors switches, only detector 1 switches, only detector 2 switches, or both switch. We determine the respective probabilities \( P_{00}, P_{01}, P_{10} \) and \( P_{11}\) by repeating the measurement many times. The individual switching probabilities of the detectors are calculated from the combined probabilities: \( P_{sw,1} = P_{10} + P_{11} \) and \( P_{sw,2} = P_{01} + P_{11} \).

The quantities \( Q_n \) and \( P_n \) are related by
\begin{equation}
  \vec{P} = M \vec{Q}
  \label{eq:PMQ}
\end{equation}
where \( \vec{P}~=~(~P_{00}~P_{01}~P_{10}~P_{11}~)^T \), \( \vec{Q}~=~(~Q_{00}~Q_{01}~Q_{10}~Q_{11}~)^T \) and
\begin{equation}
  M =
  \left( \begin{array}{rrrr}
    (1-G_1)(1-G_2) & (1-G_1)(1-E_2) & (1-E_1)(1-G_2) & (1-E_1)(1-E_2) \\
       G_2 (1-G_1) &    E_2 (1-G_1) &    G_2 (1-G_1) &    E_2 (1-E_1) \\
       G_1 (1-G_2) &    G_1 (1-E_2) &    E_1 (1-G_2) &    E_1 (1-E_2) \\
           G_1 G_2 &        G_1 E_2 &        E_1 G_2 &        E_1 E_2
  \end{array} \right)
  \label{eq:M}
\end{equation}
The parameters \( G_i \) and \( E_i \) are the switching probabilities for detector \( i \) when the corresponding qubit is in the ground state (\( G \)) or the excited state (\( E \)) respectively. Assuming no crosstalk between the detectors, these four parameters fully characterize the measurement. The calibration values are determined from independent measurements. Measuring the switching probability without applying any driving pulse provides \( G_1 = 0.507 \) and \( G_2 = 0.487 \). In another two experiments either qubit 1 or qubit 2 was resonantly excited with a pulse duration much longer than the coherence times of the qubits, so that the final state is a 50/50 incoherent mixture of the ground and excited state. The full excited state must have twice this signal. This procedure is more reliable than doing a coherent \( \pi \) rotation, since that approach is susceptible to gate errors. We measure \( E_1 = 0.760 \) and \( E_2 = 0.744 \).

The matrix \( M \) is invertible for \( G_i \neq E_i \), and the equation
\begin{equation}
  \vec{Q} = M^{-1} \vec{P}
  \label{eq:QMP}
\end{equation}
is used to extract the desired level occupations. Supplementary Figure 1 shows the data for the four measurement outcomes of the experiment described in Fig. 2.

\begin{figure}[htbp]
  \begin{center}
    \includegraphics{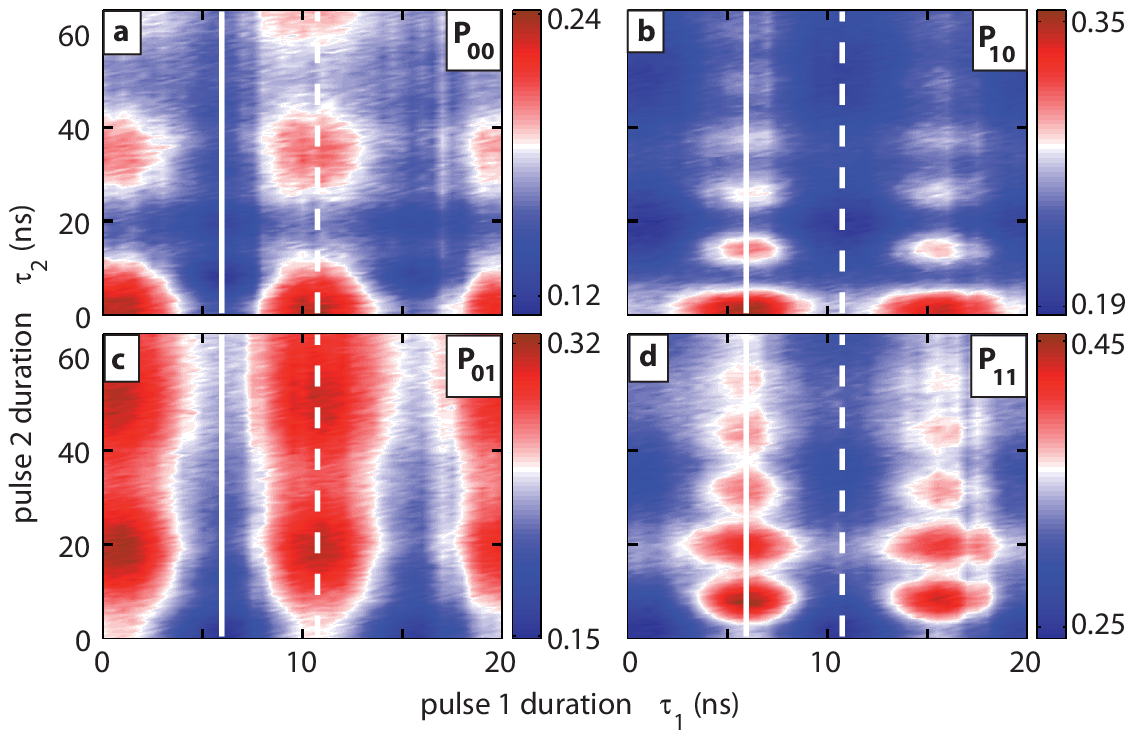}
  \end{center}
  \caption{\textbf{a}-\textbf{d} Measured combined probabilities \( P \) in color versus the durations \( \tau_1 \) and \( \tau_2 \) of pulse 1 and pulse 2.}
  \label{fig:s1}
\end{figure}

\end{document}